\documentclass[11pt]{article}
\usepackage{epsfig,times} 
\usepackage{picinpar}
\usepackage{wrapfig}
\usepackage{floatflt}
\makeatletter
\renewcommand{\section}{\@startsection{section}{1}{0in}
	{0.4\baselineskip}{0.1\baselineskip}{\Large\bf}}
\renewcommand{\subsection}{\@startsection{subsection}{2}{0in}
	{0.25\baselineskip}{-\baselineskip}{\large\bf}}
\renewcommand{\subsubsection}{\@startsection{subsubsection}{3}{0in}
	{0.1\baselineskip}{-\baselineskip}{\normalsize\bf}}
\makeatother
\begin{document}

%
\thispagestyle{myheadings}
\markright{HE 2.5.31}
\begin{center}
{\LARGE \bf EeV Hadronic Showers in Ice: The LPM effect.}
\end{center}

\begin{center}
{\bf J. Alvarez-Mu\~niz and E. Zas}\\
{\it Dept. F\'\i sica de Part\'\i culas, Universidade de Santiago, E-15706 Santiago de Compostela, SPAIN}
\end{center}

\begin{center}
{\large \bf Abstract\\}
\end{center}
\vspace{-0.5ex}

We study the longitudinal development of hadronic showers
in water and ice for energies up to 100 EeV. We present results of a
hybrid Monte Carlo method developed for the purpose of simulating
those showers.
We show parameterizations of the longitudinal development of
hadronic showers in ice and we investigate the implications of the
LPM effect on shower development. The results obtained are relevant for the
detection of high energy cosmic rays and neutrinos in
large scale detectors that use water as Cherenkov medium.

\vspace{1ex}
%
%
\section{Introduction}
\label{intro.sec}

High Energy neutrino detection is one of the inminent breakthroughs in 
particle astrophysics (AMANDA 1998). Gamma Ray Bursts, Active Galactic Nuclei
and topological defects are quite likely sites where neutrinos with energies
up to $10^{19}$ eV may be created (Halzen 1995) and in any case 
high energy neutrinos have to be produced in interactions of high energy 
cosmic rays with matter or radiation. 
Neutrinos produced in interactions of the highest energy cosmic rays with 
the galactic plane and the Cosmic Microwave Background are practically 
guaranteed. Several experiments are being developped or constructed 
using large natural volumes of water or ice as the target medium for 
neutrino interactions. Most of them are conceived to exploit the long 
range of the muons created in charged current interactions and use 
Optical Modules to detect the Cherenkov light from the muons. By looking for 
upcoming events, the atmospheric muon background can be succesfully removed.  

At high energies however the earth becomes opaque for high energy neutrinos 
and they should be expected to be arriving from the ''downgoing hemisphere'', 
that is from vertically downwards to little beyond the horizontal direction. 
To remove the background in this direction it is necessary to have energy 
resolution since the atmospheric muon rate drops rapidly with muon energy 
and provided the energy is higher than 10~PeV hardly any muons are expected. 
For this purpose the detection of the Cherenkov light from all the charged 
particles in the showers produced by high energy neutrino interactions can 
be of great help. 

Hadronic showers are mostly started by hadrons created in the fragmentation
of the nuclear debris in both charged and neutral current Deep Inelastic 
Scattering (DIS) neutrino interactions as well as in $Z^0$ resonant 
production. Most of the models of neutrino production, predict
a ratio $\nu_\mu/\nu_e\sim 2/1$ just from naive channel counting in pion 
decays. In consequence $\sim~75\%$ of the potentially
observable showers will be initiated in the hadronic vertex of the DIS 
interaction. The development of high energy hadronic showers in ice 
will play an important role in neutrino detection. 

For energies above $E_{\rm LPM}$ (2 PeV in ice) 
the cross sections for pair production and bremsstrahlung are suppressed 
due to the 
Landau-Pomeranchuk-Migdal (LPM) effect. At sufficiently high energies,
the characteristic interaction length becomes larger
than the interatomic distances and the collective atomic potentials affect
the static electric field responsible for the interaction. The LPM was 
earlier shown to strongly affect the development of electromagnetic showers 
in ice above 20 PeV (Alvarez 1997). Electromagnetic showers of those energies 
are significantly deviated from the Greisen parameterizations with
an increase in shower length\footnote{Defined as the 
length along which shower size exceeds
$70\%$ of its maximum.} with shower energy $E_0$ which
can be approximately parameterized as $E_0^{1/3}$. 
In this work we study the influence of the
LPM effect on the development of
hadronic showers initiated by neutrino interactions in ice. 
Our results are also relevant for the 
calculation of the coherent radio pulses generated by the excess 
negative charge developed by showers in dense media, but this will be 
discussed in a separate presentation (Alvarez 1999).  

\section{Hadronic showers: Simulations and results}
\label{simulations.sec}

An hadronic shower can be thought of having a penetrating core containing
mostly pions
which continually produces electromagnetic subshowers fundamentally 
through $\pi^0$ decay in two photons. In dense media 
charged pions are expected to interact before decaying.
In a simple model of shower development, we can consider that 
after a neutrino interaction the energy of the produced hadrons is 
roughly the energy transferred to the nucleus ($yE_\nu$ where $E_\nu$ is the
neutrino energy in the LAB system), divided by the multiplicity of the 
interaction, assuming equipartition
of energy. For a 1 EeV energy transfer to the hadron, the average multiplicity 
is $\sim$ 14 and the average energy of the photons produced by $\pi^0$ decays
will be of the order of 35 PeV. These photons will initiate 
electromagnetic subshowers which are expected to be
little affected by the LPM effect (Alvarez 1997).
One can naively think that as $yE_\nu$ increases and so does the photon energy
the electromagnetic subshowers will exhibit strong LPM behaviour with the
typical elongations in their longitudinal development. 
In ice however this is not the case
since $\pi^0$ interaction dominates over decay for energies above 
about 6.7 PeV. It is then expected that even 100 EeV hadronic showers initiated 
in neutrino interactions will show LPM effects in a mitigated form.

\begin{figwindow}[1,r,%
{\mbox{\epsfig{file=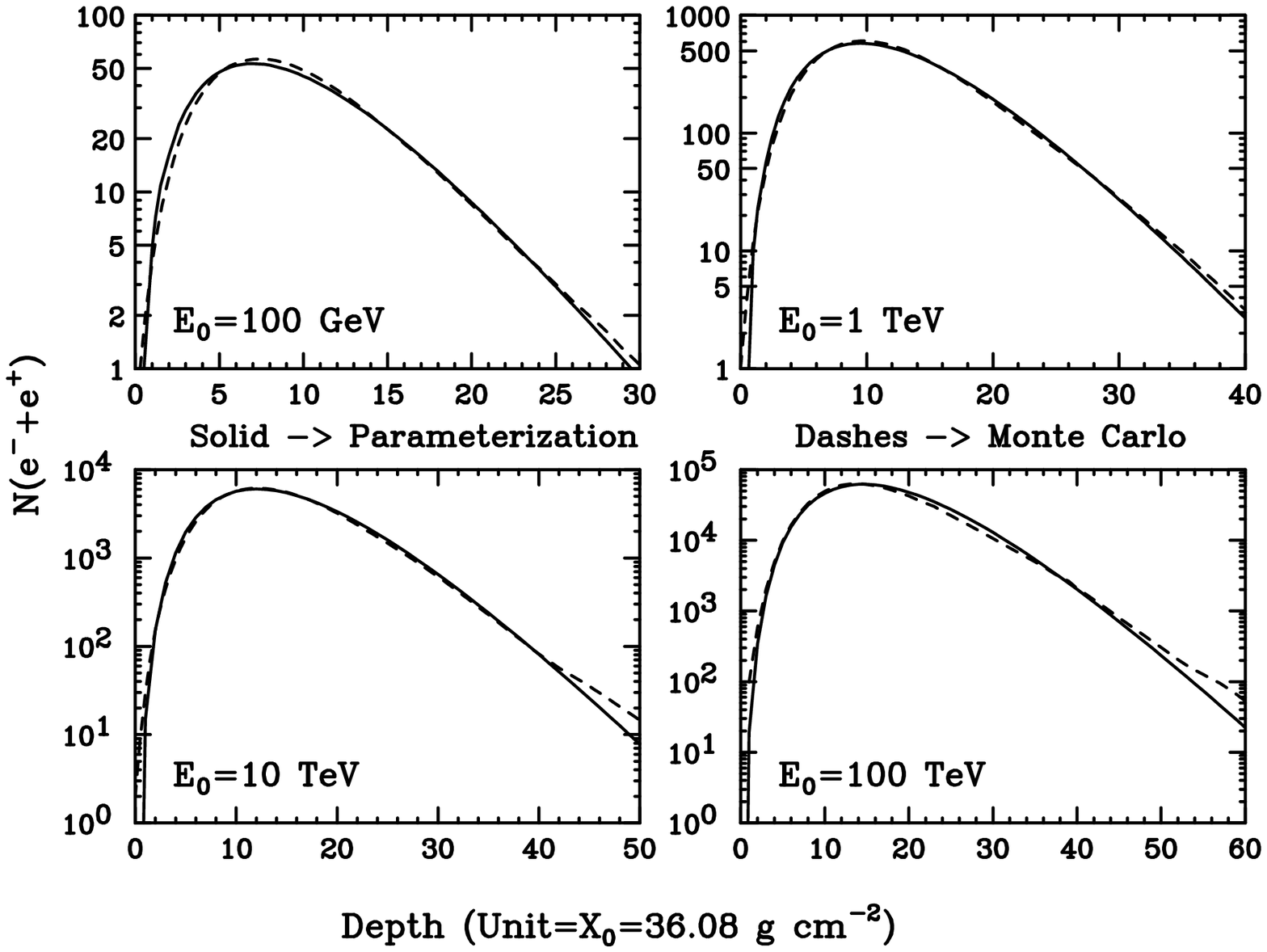,width=5.0in}}},%
{Longitudinal development of proton showers in ice.
The solid line represents
the fit to a Monte Carlo (dashed lines).}]
To confirm these interpretations we have simulated showers initiated in
the hadronic vertex of a neutrino interaction. We have modelled the hadronic 
shower using the average multiplicity and the distribution of momentum of
the secondaries measured in HERA (ZEUS 1999). 
For this purpose we have developed a fast hybrid Monte Carlo
which simulates 1 dimensional showers down to a certain crossover energy,
at which the subshowers produced are taken from tested parameterizations.
For the purely electromagnetic subshowers we use the Greisen and NKG 
parameterizations. For the hadronic showers initiated by low energy 
protons, pions or kaons we have developed our own 
set of parameterizations which are valid for energies below 100 TeV. The
parameterizations together with the results of the simulations for proton
showers are shown in Fig.~1. The longitudinal development has been fitted 
to a function similar to the one used in (Gaisser 1990) of the form: 
\end{figwindow}

\begin{equation}
N(t)=S_0 {E_0\over E_c} \left( {X_{\rm max}-\lambda\over X_{\rm max}} \right)
e^{{X_{\rm max}\over \lambda} - 1}
\left({t\over X_{\rm max}-\lambda}\right)^{X_{\rm max}/ \lambda}
e^{-t/ \lambda}~,
\label{hadronicparam}
\end{equation}
where $X_{\rm max}=X_0~{\rm log} {E_0\over E_c}$ and 
$S_0$, $X_0$, $\lambda$ y $E_c$ are free parameters. 
Their values for different types of primaries,
are shown in table I.

\begin{center}
\begin{tabular}{c|ccc} \hline
Primary & Proton & Pion & Kaon \\ \hline
$S_0$& 0.11842 & 0.036684 & 0.0298 \\
$X_0~({\rm g~cm}^{-2})$ & 39.562 & 36.862 & 36.997 \\
$\lambda~({\rm g~cm}^{-2})$& 113.03 & 115.26 & 119.61 \\
$E_c~({\rm GeV})$& 0.17006  & 0.052507 & 0.048507 \\ \hline
\end{tabular}
\end{center}
\begin{center}
{{\it {\bf Table I:} Values of the parameters for the fit to 
expression \ref{hadronicparam} of the 
longitudinal development of hadronic showers in ice.}}
\end{center}

We have studied hadronic showers initiated in neutrino interactions in ice
for energies up to 100 EeV.
We have chosen to simulate several quantities that are relevant for Cherenkov
emission such as   
the fraction of energy going into electromagnetic subshowers, which is seen to 
increase with shower energy reaching values as high as $94\%$ at EeV energies
(Alvarez 1998), 
and the total and excess charge tracklengths, 
which are respectively dominated by the contribution 
of electrons and positrons and by the excess of electrons over positrons.
For the latter purpose we have used parameterizations obtained in (Zas 1993). 
\begin{figwindow}[1,r,%
{\mbox{\epsfig{file=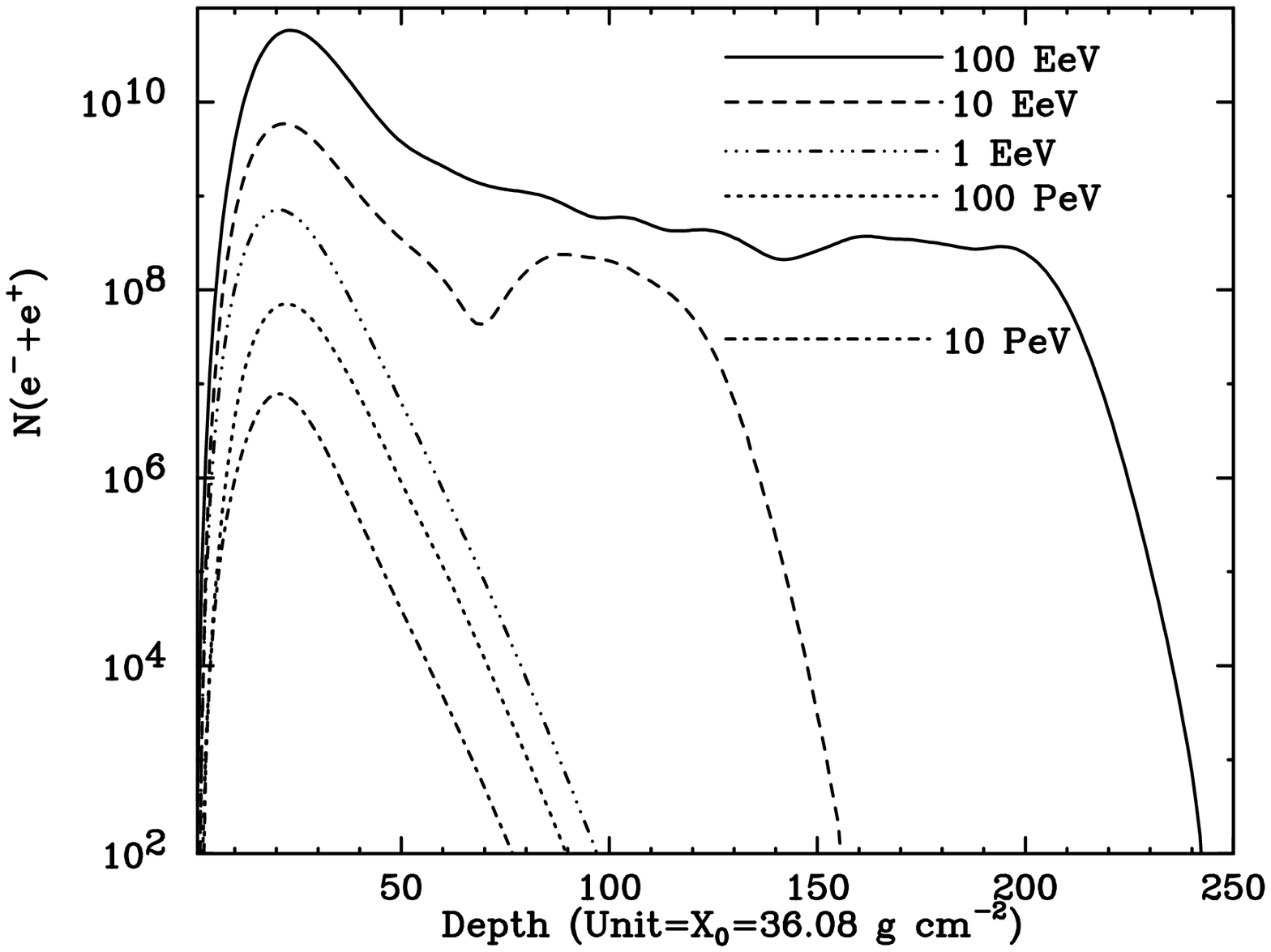,width=5.0in}}},%
{Longitudinal development of hadronic showers initiated by neutrino
interactions in ice. }]
In Fig.~2 we show the longitudinal development of hadronic
showers. Below 1 EeV the longitudinal development "scales" with shower energy
and it is not affected by the LPM effect in agreement with the
interpretation given above. 
This is not surprising. Due to the high multiplicities involved in
hadronic interactions
the energy of the $\pi^0$'s is considerably reduced with respect to the
primary energy. The average energy of the $\pi^0$'s (as produced by SIBYLL) 
in a proton-proton collision at
$10^{19}$ eV in the LAB frame is of the order of 17 PeV. 
Moreover we have obtained that only about $10\%$ of 
the $\pi^0$'s of energy greater than 20 PeV, produced in proton-ice 
interactions, are expected to 
decay in ice producing photons of energy above 20 PeV. As a conclusion
showers are not elongated despite being produced by primaries with
energies well above $E_{\rm LPM}$.
\end{figwindow}
 
We have found that a fraction of showers above 
1 EeV have deep tails characteristic of LPM showers. These tails are produced
by the electromagnetic decays of resonances with short lifetimes that are
created in early interactions in the shower. In particular we found that the
$\eta$ and $\eta'$ contribute most to this effect. 
Although the result is model dependent,
it is very interesting since if these showers are ever observed, they would
provide experimental information on the production of resonances and their
decays in electromagnetic particles.

\begin{figwindow}[1,r,%
{\mbox{\epsfig{file=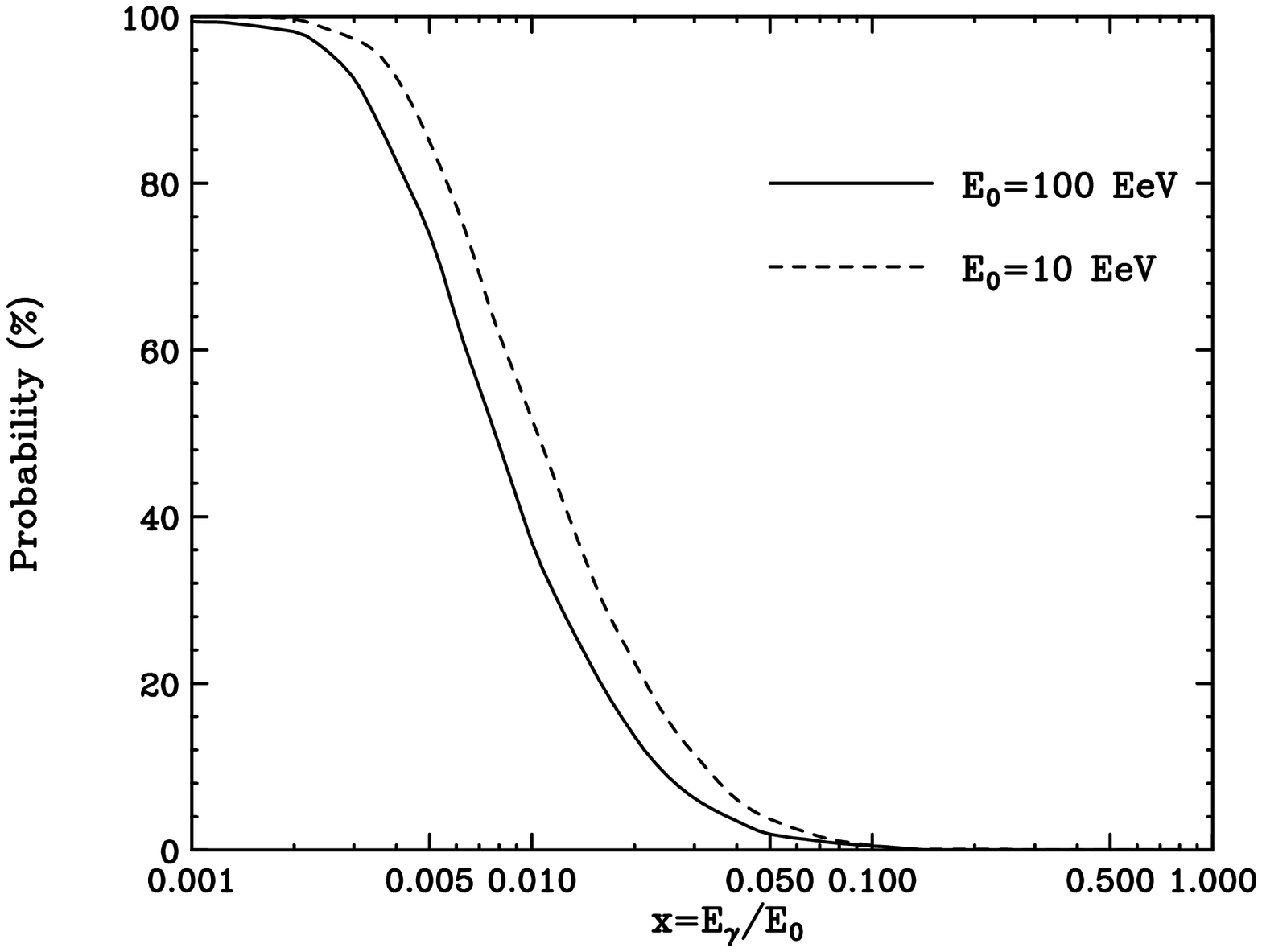,width=5.0in}}},%
{Probability of having a photon of energy greater than $E_\gamma=xE_0$ 
in a neutrino hadronic shower of energy $E_0$.}]
The probability of having a neutrino hadronic shower with an LPM tail can be
computed with the aid of Figure 3, 
in which we plot the integral energy distribution of 
the most energetic photon obtained in showers initiated in the 
hadronic vertex of a neutrino interaction. The plot represents the 
probability of having a photon with a energy 
($E_\gamma=xE_0$) greater than $xE_0$ where $E_0$ is the shower energy. 
From the plot we can 
see that 10 EeV showers have a probability of $50\%$ of having a photon
with energy greater than 100 PeV which will produce a long LPM tail. 
Approximately the same distribution is obtained for a 100 EeV shower, 
from which it can be deduced that it is practically impossible to have 
a 100 EeV shower without LPM effect. 

In summary, we have shown that hadronic showers produced in neutrino 
interactions are very different from electromagnetic showers being
much less affected by the LPM effect. We can expect the hadronic showers 
induced by neutrinos of energy below 1 EeV/$y$ where $y$ is the fraction of 
energy transferred to the hadron to have a quite ordinary 
longitudinal development without the typical LPM tails. 
Our results are relevant for radio emission from hadronic showers 
which is treated in (Alvarez 1999)

\end{figwindow}

{\bf Acknowledgements:} We thank T. Stanev for making
UNICAS available to us and for many discussions
and suggestions on this work, we also thank
R.A. V\'azquez for useful comments and for carefully
reading the manuscript and L.G.~Dedenko, F. Halzen, J. Knapp,
and P.~Lipari for fruitful discussions. This work was supported in part
by CICYT (AEN96-1673) and by Xunta de Galicia (XUGA-20602B98). One of
the authors J. A. thanks the Xunta de Galicia for financial support.

\vspace{1ex}
\begin{center}
{\Large\bf References}
\end{center}
Alvarez-Mu\~niz, J., V\'azquez, R.A. \& Zas, E, 1999 
These proceedings (HE 6.13.14)\\
Alvarez-Mu\~niz, J., \& Zas, E., 1997, Phys. Lett. B 411, p.218\\
Alvarez-Mu\~niz, J., \& Zas, E., 1998, Phys. Lett. B 434, p.396\\
AMANDA Collaboration, 1998, $18^{\rm th}$ Int. Conf. on Neutr. Phys. and 
Astroph. (Neutrino 98) Japan, hep-ex/9809025\\
Gaisser, T.K., 1990, {\it Cosmic Rays and Particle Physics}, 
Cambridge University Press\\
Halzen, F., Gaisser, T. \& Stanev, T., 1995, Phys. Rep. 258, p.173 \\
Zas, E., Halzen, F., Stanev, T., 1992, Phys. Rev. D 45, 1, p.362\\
ZEUS Collaboration, 1999, hep-ex/9903056 \\
\end{document}